\documentclass[letterpaper,twocolumn,10pt]{article}

\usepackage{zhanggroup}  
\usepackage{times}  
\usepackage{helvet} 
\usepackage{courier}  
\usepackage{graphicx}  
\usepackage{caption} 
\usepackage{color}
\usepackage{xcolor}
\usepackage{xspace}
\usepackage[font=footnotesize]{subfig}
\usepackage{booktabs} 
\usepackage{multirow}

\newcommand{\mypara}[1]{\bigskip\noindent\textbf{#1.} \xspace}

\begin{document}
\date{}

\author{
Yang Zou\thanks{The first two authors made equal contributions.}\ \ \
Zhikun Zhang\textsuperscript{$\ast$}\ \ \
Michael Backes\ \ \
Yang Zhang
\\
\\
\textit{CISPA Helmholtz Center for Information Security} \\
}

\title{Privacy Analysis of Deep Learning in the Wild: \\ Membership Inference Attacks against Transfer Learning}
\maketitle

\begin{abstract}
While being deployed in many critical applications as core components, machine learning (ML) models are vulnerable to various security and privacy attacks. 
One major privacy attack in this domain is membership inference, where an adversary aims to determine whether a target data sample is part of the training set of a target ML model. 
So far, most of the current membership inference attacks are evaluated against ML models trained from scratch. 
However, real-world ML models are typically trained following the transfer learning paradigm, where a model owner takes a pretrained model learned from a different dataset, namely teacher model, and trains her own student model by fine-tuning the teacher model with her own data.

In this paper, we perform the first systematic evaluation of membership inference attacks against transfer learning models. 
We adopt the strategy of shadow model training to derive the data for training our membership inference classifier. 
Extensive experiments on four real-world image datasets show that membership inference can achieve effective performance. 
For instance, on the CIFAR100 classifier transferred from ResNet20 (pretrained with Caltech101), our membership inference achieves $95\%$ attack AUC.
Moreover, we show that membership inference is still effective when the architecture of target model is unknown. 
Our results shed light on the severity of membership risks stemming from machine learning models in practice.

\end{abstract}

\section{Introduction}
\label{section:introduction}

Machine learning (ML) is powerful to provide automated extraction of insights from data utilizing predictive models. 
Thus, it has been deployed in a variety of critical applications ranging from autonomous systems~\cite{VL14,LMSC17} to advertising~\cite{YYMRHL18}.
However, previous studies have shown that the ML models are vulnerable to various privacy attacks, such as adversarial example~\cite{TKPGBM17,CW17}, model inversion~\cite{FLJLPR14,FJR15}, model stealing~\cite{TZJRR16,OSF19} and membership inference~\cite{SSSS17,SZHBFB19,CYZF20}.

In this paper, we focus on membership inference attack, where the adversaries' goal is to determine whether a specific data sample is used to train the ML models.
Knowing the membership status of individual user's data may cause severe information leakage.
For example, knowing that a certain patient's clinical records were used to train a model associated with a disease (e.g., to determine the appropriate drug dosage or to discover the genetic basis of the disease) can reveal that the patient carries the associated disease.

Previous studies~\cite{LBWBWTGC18,NSH18,SZHBFB19,YGFJ18,CTCP20,LZ20} on membership inference mainly target on the models trained from scratch.
However, training ML models following this paradigm requires large amount of data and computational resources, which is infeasible for individual researchers and small companies.
In practice, most of the ML models are trained following the \emph{transfer learning} paradigm.
Transfer learning is an effective approach to transfer the ``knowledge'' learned from a pretrained model (teacher model) to a new model (student model)~\cite{WYVZZ18}.
There are two approaches for transferring from teacher model to student model.
One is feature extractor that freeze the first $K$ layers of teacher model and train the output layers of the student model.
The other is fine-tuning that use the parameters of teacher model as initialization and fine-tune the model with student model's datasets.
Transfer learning enables individual researchers and small companies to effectively train an ML with good performance.

\subsection{Our Contributions}
In this paper, we take the first step to investigate the effectiveness of membership inference against transfer learning.
Different from classical membership inference, there are two attack surfaces in the transfer learning setting, membership status of training data of both student model and teacher model.
Besides, the adversaries have access to either student model or teacher model according to their capabilities.
In total, there are three different attacks in the transfer learning setting.

Similar to previous membership inference attack, we assume the adversaries only have black-box access to the student model or teacher model, which is the most difficult scenarios for the adversaries.
We adopt the strategy of shadow model training to generate the data for training the attack models.
In this paper, we seek to answer the following questions:
\begin{itemize}
    \item Whether the three membership inference attacks is still effective in the transfer learning setting?
    \item What is the impact of frozen layers on the effectiveness of membership inference?
    \item Whether the membership inference is still effective when the architecture of teacher model is unknown?
\end{itemize}

We systematically design the attack pipeline in the transfer learning setting and conduct extensive experiments on multiple real-world datasets to answer the above questions. 
The experimental results show that membership inference attacks against teacher and student model are still effective with attack AUC larger than $0.9$.
When the model is transferred from teacher model to student model, the membership information of teacher model will not be leaked.
Besides, we observe that the performance of membership inference in the student model significantly decrease when the number of frozen layers increase.
Finally, we demonstrate that the membership inference is still effective when the target model and shadow model are trained by different model architectures.

The contributions of this paper are three-fold:

\begin{itemize}
    \item We take the first step to investigate the membership privacy against transfer learning.
    
    \item We systematically design a novel attack pipeline for the three membership inference attacks in the transfer learning setting.

    \item We conduct extensive experiments to illustrate the effectiveness of three different attacks.
    We also show the impact of the number of frozen layers, and the effectiveness of the attack when the architecture of the teacher model is unknown.
\end{itemize}

\section{Related Work}

\subsection{Membership Inference}
Membership inference attacks have been extensively studied in many different data domains, ranging from biomedical data~\cite{BBHM16, HZHBTWB19} to mobility traces~\cite{PTC18}.
Shokri et al.~\cite{SSSS17} presented the first membership inference attack against ML models. 
The general idea is to use shadow models to mimic the target model’s behavior to generate training data for the attack model.
Salem et al.~\cite{SZHBFB19} gradually removed the assumptions of~\cite{SSSS17} by proposing three different attacks.
Since then, membership inference has been extensively investigated in various ML models, such as federated learning~\cite{MSCS19}, generative adversarial networks~\cite{HMDC19, CYZF20}, natural language processing~\cite{SS19}, and computer vision segmentation~\cite{HRSF20}.
To mitigate the threat of membership inference, a number of defense mechanisms have been proposed, which can be classified into three categories: 
reducing overfitting~\cite{SSSS17,SZHBFB19,LLR20}, perturbing posteriors~\cite{JSBZG19}, and adversarial training~\cite{NSH18}.

However, all of the previous studies focus on the target models training from scratch.
This is the first work studying membership inference in the practical transfer learning setting.

\subsection{Attacks against Transfer Learning}
Previous studies have shown that transfer learning is vulnerable to adversarial attacks~\cite{WYVZZ18} and data poisoning attacks~\cite{SSMS20}.
Wang et al.~\cite{WYVZZ18} propose an adversarial attack against transfer learning.
The general idea is to optimize an image to mimic the internal representation of the target image, instead of optimizing the image to be predicted as the target label as classical adversarial attacks~\cite{CW17}. 
Schuster et al.~\cite{SSMS20} propose a data poisoning attack against word embedding.
Embedding based natural language processing tasks follow the transfer learning paradigm, where the embedding model and downstream model can be regarded as teacher model and student model, respectively.
The objective is to either make the target word ranks higher among others, or to move the target word closer/farther from a particular set of words.
The authors conduct extensive experiments to show that the attack on the embedding model (teacher model) can severely influence multiple downstream models (student models).

These previous studies target at misleading transfer learning models' behavior; 
while this paper targets at inferring the private information of the training dataset of the transfer learning models.

\section{Background}
\label{section:background}

In this section, we introduce some background knowledge about transfer learning and membership inference attack.

\subsection{Transfer Learning}

To achieve higher performance, the deep learning models have become more and more complex.
However, training these complex models from scratch requires large amount of training data and computational resources.
For example, the OpenAI company spends  almost $1.4$ TB training data and $4.6$ million USD to train the GPT-3 model.\footnote{\url{https://towardsdatascience.com/gpt-3-primer-67bc2d821a00}}
Oftentimes, individual researchers and small companies cannot afford such amount of resources.
To address this issue, the researchers propose the transfer learning paradigm.
The general idea is to let the leading companies collect large amount of data from their users and pretrain the complex deep learning models with plenty of computational resources; then, the pretrained models are published for downstream applications.
Then, individual researchers and small companies no longer need to train their complex deep learning models from scratch.
Instead, they can use the pretrained model as starting point and customize their own applications using their local datasets.

Typically, in transfer learning, the pretrained model is referred to as teacher model, and the downstream model is referred to as student model.
\begin{figure}[!t]
\centering 
\includegraphics[width=0.45\textwidth]{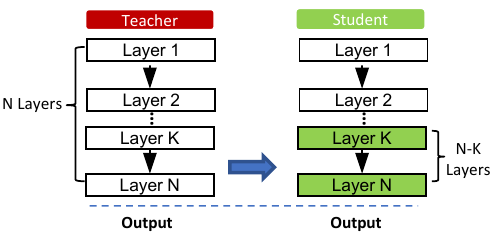}
\caption{Workflow of transfer learning. For the feature extractor approach, the student model is initialized by copying the first $K$ layers from teacher model, and updated for the last $N-K$ layers.
For the fine-tuning approach, the student model is updated for the whole layers.
}
\label{figure:workflow_of_TL}
\end{figure}

\subsubsection{Workflow of Transfer Learning.}
The workflow of transfer learning is illustrated in Figure~\ref{figure:workflow_of_TL}.
In practice, there are two kinds of transfer learning approaches:
\begin{itemize}
    \item \mypara{Feature Extractor}
    The student model freeze the first $K$ layers of teacher model, where $K \leq N - 1$, and treat them as a feature extractor. 
    Then, the student model use back propagation to update the parameters of the rest of layers using their local datasets.
    This approach is suitable for the ML models with huge amount of layers and parameters, such as GPT-3 with $175$ billion parameters.
    
    \item \mypara{Fine-tuning}
    Instead of freezing part of the layers of the teacher model, the fine-tuning approach aims to update the parameters of the whole ML model.
    Since the parameters of the teacher model are well-trained, the student model can converge fastly and achieve high performance~\cite{EMBBV09}.
\end{itemize}

\subsection{Membership Inference against ML Models}

In machine learning, the objective of membership inference is to determine whether a data sample was used to train the machine learning models.
Knowing the membership status of individual user's data may cause severe information leakage.
For example, knowing that a certain patient's clinical records were used to train a model associated with a disease (e.g., to determine the appropriate drug dosage or to discover the genetic basis of the disease) can reveal that the patient carries the associated disease.

Formally, membership inference attack can be defined as the following function:
$$
\mathcal{A}: \textbf{x}_{target}, \mathcal{M}, \mathcal{K} \rightarrow \{ 0, 1 \}
$$
Here, $\textbf{x}_{target}$ is a target data sample, $\mathcal{M}$ is the target model, and $\mathcal{K}$ is the auxiliary knowledge of adversaries.
The output value equals $1$ means that $\textbf{x}_{target}$ is a member of $\mathcal{M}$’s training dataset $\mathcal{D}_{Train}$ and $0$ otherwise. 
The attack model $\mathcal{A}$ is essentially a binary classifier.

\section{Attack Methodologies}
\label{section:methodology}

In this section, we first introduce the attack taxonomy and threat model.
Then, we depict the pipeline of our attacks.

\subsection{Attack Taxonomy and Threat Model}

Different from the classical membership inference, there are two attack surfaces in the transfer learning setting:
1) determine whether a data sample is used to train the teacher model; 
2) determine whether a data sample is used to train the student model.
Besides, based on the capabilities of different adversaries, they may have access to either teacher model or student model.
Considering two different attack surfaces and two different adversary capabilities, there are in total three types of attacks:

\begin{itemize}
    \item \mypara{Attack-1}
    The adversaries have access to the teacher model and aims to infer the membership status of teacher dataset.
    This type of attack is similar to the classical membership inference attack where the target model is trained from scratch.
    
    \item \mypara{Attack-2} 
    The adversaries have access to the student model and aims to infer the membership status of teacher dataset.
    In this attack, the target model is not directly trained from the target dataset.
    
    \item \mypara{Attack-3}
    The adversaries have access to the student model and aims to infer the membership status of student dataset.
    Different from Attack-1, the target model of this attack is transferred from the teacher model.
\end{itemize}

Notice that another combination that the adversaries have access to the teacher model and aims to infer the membership status of student model is infeasible, since the student model is never used for training the teacher model.

Similar to previous membership inference attacks~\cite{SSSS17, SZHBFB19}, we assume the adversares have black-box access to either the teacher model and student model.
This means that the adversaries can only query these models and obtain their corresponding posteriors.
Compared to the white-box setting, where the adversaries have direct access to the architecture and parameters of the target model, the black-box setting is more realistic, and more challenging for the adversaries~\cite{NSH19}.
We further assume that the adversaries have a shadow dataset which can be used to train a set of shadow models to mimic the behavior of the target model.
The shadow models are then used to generate another dataset to train the attack model.
We will relax this assumption in the evaluation part.

\begin{figure*}[!t]
\centering  %
\includegraphics[width = 0.75\textwidth]{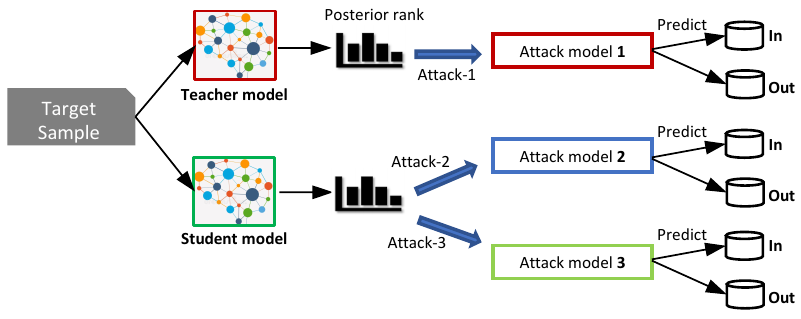}
\caption{Overview of the attack pipeline.
Given a target data sample, the adversaries first query the target model (teacher/student) and obtain the corresponding posterior.
The adversaries feed the posterior to the attack model to determine the membership status.
``In'' and ``Out'' stand for member and non-member, respectively.
}
\label{figure:attack_phase}
\end{figure*}

\subsection{Attack Pipeline}
The general attack pipeline is illustrated in Figure~\ref{figure:attack_phase}.
Given a data sample and the objective (determine membership status of teacher dataset or student dataset), the adversaries first query the corresponding target model (teacher model or student model) and obtain the posterior.
Then, the obtained posterior is fed to the corresponding attack model for determining the membership status.

\begin{figure}[!t]
\centering %
\includegraphics[width = 8.3cm]{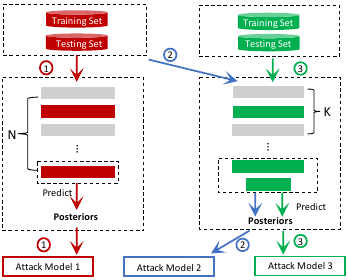}
\caption{Overview of the attack models training.
The middle left part and middle right part stand for the teacher shadow model and student shadow model, respectively.
The outputs of these two shadow models are the posteriors used to train the attack models.
Notice that the two shadow models are slightly different in the last few layers, since their prediction tasks are different.
}
\label{figure:train_phase}
\end{figure}

\mypara{Attack Models Training}
Given the attack pipeline, the next question is how to train the three attack models.
Recall that the attack model is essentially a binary classifier that takes as input the posteriors of the target model and determine the membership status of the target sample.
We adopt the strategy of shadow model training to derive the data for training our attack classifier.
For ease of presentation, we assume that the adversaries have access to local datasets that come from the same distribution of either teacher dataset or student dataset.
We will relax this assumption in the evaluation part.
The training processes of three attacks are illustrated in Figure~\ref{figure:train_phase}.

\section{Evaluation}
\label{section:experiments}
In this section, we conduct empirical experiments on multiple real-world datasets to illustrate the effectiveness of membership inference against transfer learning.
Specifically, we aim to answer the following key questions:

\begin{itemize}
    \item \mypara{Q1} Whether the three membership inference attacks is still effective in the transfer learning setting?
    
    \item \mypara{Q2} What is the impact of frozen layers on the effectiveness of membership inference?
    
    \item \mypara{Q3} Whether the membership inference is still effective when the architecture of teacher model is unknown?
\end{itemize}

\subsection{Experimental Setup}

\mypara{Environment}
Our experiments are conducted on a server with Intel Xeon E7-8867 v3 @ 2.50GHz and 2.5TB memory and implement the attacks using Keras.

\mypara{Datasets} In our evaluation, we use the following image datasets.

\begin{itemize}
    \item \textbf{Caltech101.}
    It contains $5,486$ training images and $3,658$ testing images, which split between $101$ distinct object categories (faces, watches, ants, pianos, etc.) and a background category.
    Specially, there are about $40$ to $800$ images in each category and most of the categories have about $50$ images.
    
    \item \textbf{CIFAR100.}
    It is a benchmark dataset widely used to evaluate the image recognition algorithms.
    The dataset is composed of color images equally classified into $100$ classes such as food, people, insects, etc.
    For each class, there are $500$ training images and $100$ testing images.
    
    \item \textbf{Flowers102.}
    This dataset consists of $102$ flower categories, which are commonly occurring in UK.
    It contains $6,149$ training images and $1,020$ testing images spreading in $102$ classes.
    Notice that the Flowers$102$ dataset is unbalanced.
    
    \item \textbf{PubFig83.}
    This dataset consists of $8,300$ cropped facial images coming from $83$ public facial images, each contains $100$ variants.
    The images in PubFig$83$ are taken from the web and not collected in a controlled environment.
\end{itemize}

\mypara{Metrics} 
In our experiments, we use four metrics, \textbf{accuracy}, \textbf{precision}, \textbf{recall} and \textbf{AUC}, to measure the attack performance. 
In particular, \textbf{AUC} is a widely used metric to measure the performance of binary classification for a number of thresholds~\cite{FLJLPR14,BHPZ17,PZ17,PZ172,PTC18,HZHBTWB19,SZHBFB19,Z19,JSBZG19,ZHSMVB20}.
Typically, higher AUC value implies better ability to predict the membership status.

\mypara{Experimental Settings}
In our experiments, we use the Caltech101 dataset to train the teacher model.
The other three datasets are used to train the student model.

For each dataset $\mathcal{D}$, we evenly split it into two disjoint parts, the target dataset $\mathcal{D}_{target}$ and the shadow dataset $\mathcal{D}_{shadow}$.
We further split $\mathcal{D}_{shadow}$ into shadow training dataset $\mathcal{D}_{shadow}^{train}$ and shadow testing dataset $\mathcal{D}_{shadow}^{test}$.
The $\mathcal{D}_{shadow}^{train}$ and $\mathcal{D}_{shadow}^{test}$ datasets are used to generate the ground-truth member and non-member data for training the attack model.
The same procedure applies to the target dataset $\mathcal{D}_{target}$, resulting in $\mathcal{D}_{target}^{train}$ and $\mathcal{D}_{target}^{test}$.
The target datasets are used to evaluate the attack performance.

By default, we set the number of training epoch to $50$, the number of shadow models to $1$, and the learning rate to $0.001$.
We use ResNet20~\cite{HZRS16} as the target model.

\subsection{Q1: Effectiveness of Different Membership Inference Attacks}
In this subsection, we evaluate the performance of three membership inference attacks on three different student datasets.
We use the fine-tuning approach for transfer learning.
Table~\ref{table:result_of_attack_1},~\ref{table:result_of_attack_2} and~\ref{table:result_of_attack_3} illustrate the attack performance of the three attacks, respectively.

\begin{table}[!tpb]
\centering
\caption{Attack performance of Attack-1.
The teacher model is trained on Caltech101 dataset.}
\label{table:result_of_attack_1}
\begin{tabular}{c  c  c  c  c}
\toprule
Dataset & AUC & Accuracy & Precision & Recall\\
\midrule
 Caltech101 & 0.906 & 0.753 & 0.644 & 0.882 \\
\bottomrule
\end{tabular}
\end{table}

\begin{table}[!tbp]
\centering
\caption{Attack performance of Attack-2.
The target models are first pretrained on Caltech101 dataset and fine-tuned on three different datasets.}
\label{table:result_of_attack_2}
\begin{tabular}{ c  c  c  c  c}
\toprule
Dataset & AUC & Accuracy & Precision & Recall\\
\midrule
CIFAR100 & 0.522 & 0.502 & 0.478 & 0.523 \\
Flowers102 & 0.528 & 0.496 & 0.432 & 0.505 \\
PubFig83 & 0.495 & 0.481 & 0.396 & 0.524 \\
\bottomrule
\end{tabular}
\end{table}

\begin{table}[!tpb]
\centering
\caption{Attack performance of Attack-3.
The target models are first pretrained on Caltech101 dataset and fine-tuned on three different datasets.}
\label{table:result_of_attack_3}
\begin{tabular}{c  c  c  c  c}
\toprule
Dataset & AUC & Accuracy & Precision & Recall\\
\midrule
 CIFAR100 & 0.954 & 0.850 & 0.779 & 1.0 \\
 Flowers102 & 0.994 & 0.897 & 0.834 & 0.959 \\
 PubFig83 & 0.998 & 0.935 & 0.887 & 0.975 \\
\bottomrule
\end{tabular}
\end{table}

\mypara{Attack-1}
To recap, the adversaries have access to the \emph{teacher model}, and aim to determine the membership status of \emph{teacher dataset}, which is similar to the classical membership inference.
In this experiment, we perform experiment on Caltech101 dataset.
Table~\ref{table:result_of_attack_1} illustrates the attack performance.
The experimental results show that the adversaries can achieve $0.753$ accuracy, $0.644$ precision, $0.882$ recall, and $0.906$ AUC.
This result is consistent with that of~\cite{SZHBFB19}. 

\mypara{Attack-2}
Recall that in this attack the adversaries only have access to the \emph{student model} and aim to determine the membership status of \emph{teacher dataset}.
We first train the teacher model on Caltech101 dataset, and fine-tune the teacher model on three student datasets (CIFAR$100$, Flowers$102$ and PubFig$83$) to obtain the student models.
We feed the teacher dataset Caltech101 to the student model and obtain the training samples for the attack model.
Table~\ref{table:result_of_attack_2} illustrates the attack performance on three student datasets.
We observe that the attack AUC and recall are close to $0.5$, which is similar to random guessing.
Comparing the attack performance of Attack-1 and Attack-2, we reveal that the membership information of the teacher dataset is disappeared when the teacher models are fine-tuned by the student dataset.
In another word, the student model would not leak membership information of the teacher dataset.

\mypara{Attack-3}
The objective of this attack is to determine the membership status of \emph{student dataset} when the adversaries have access to the \emph{student model}.
Similar to Attack-2, we first train the teacher model on Caltech101 dataset, and fine-tune the teacher model on other three datasets to obtain the student models.
The main difference is that in this attack we feed the student datasets to the student models to obtain the training data for attack model.
Table~\ref{table:result_of_attack_3} summarizes the attack performance on three student models.
We observe that the membership inference performs well on three student models, meaning the membership inference is still effective in student models.

\begin{figure*}[!ht]
\centering
\includegraphics[width=0.46\textwidth]{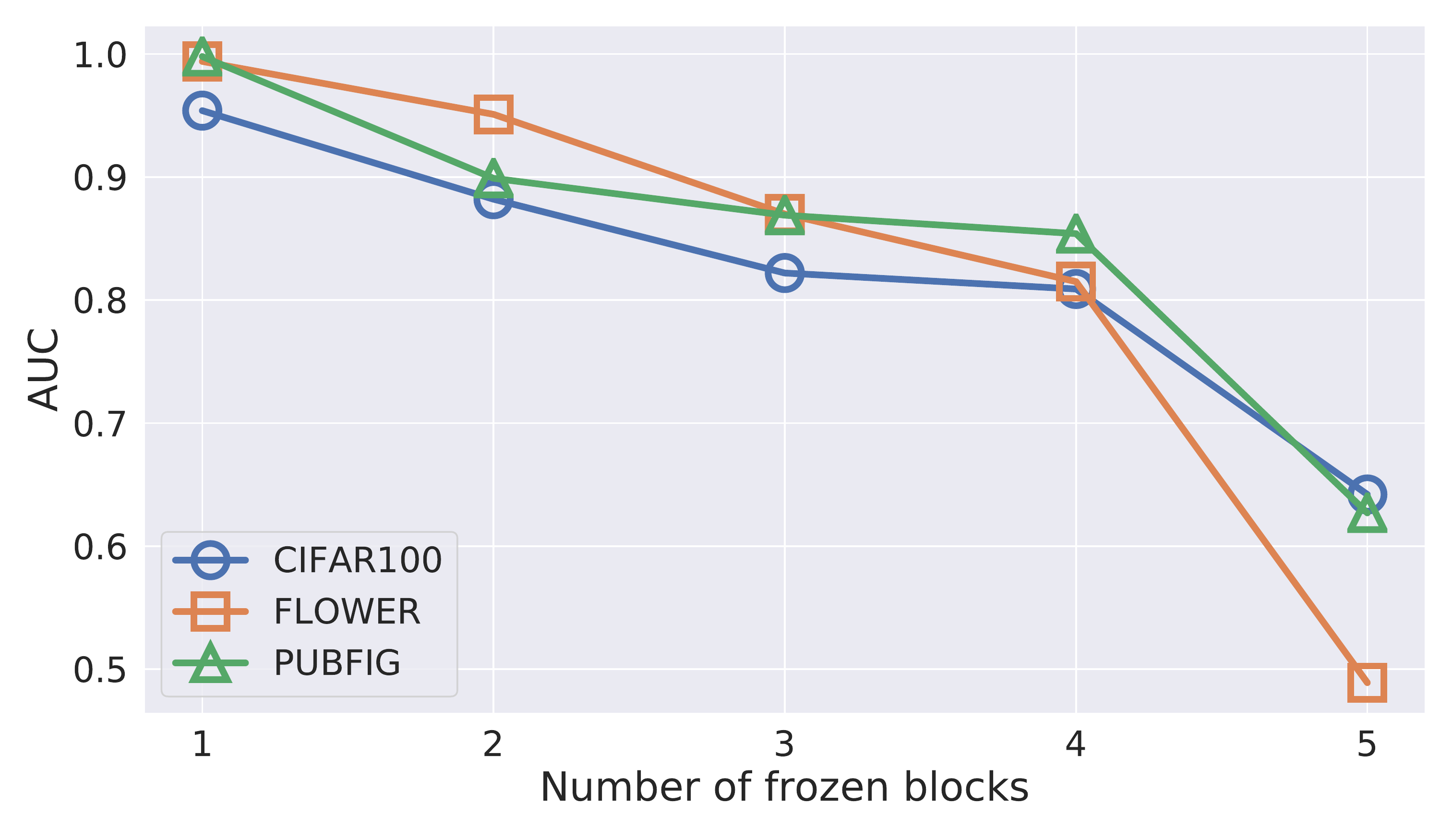}
\includegraphics[width=0.46\textwidth]{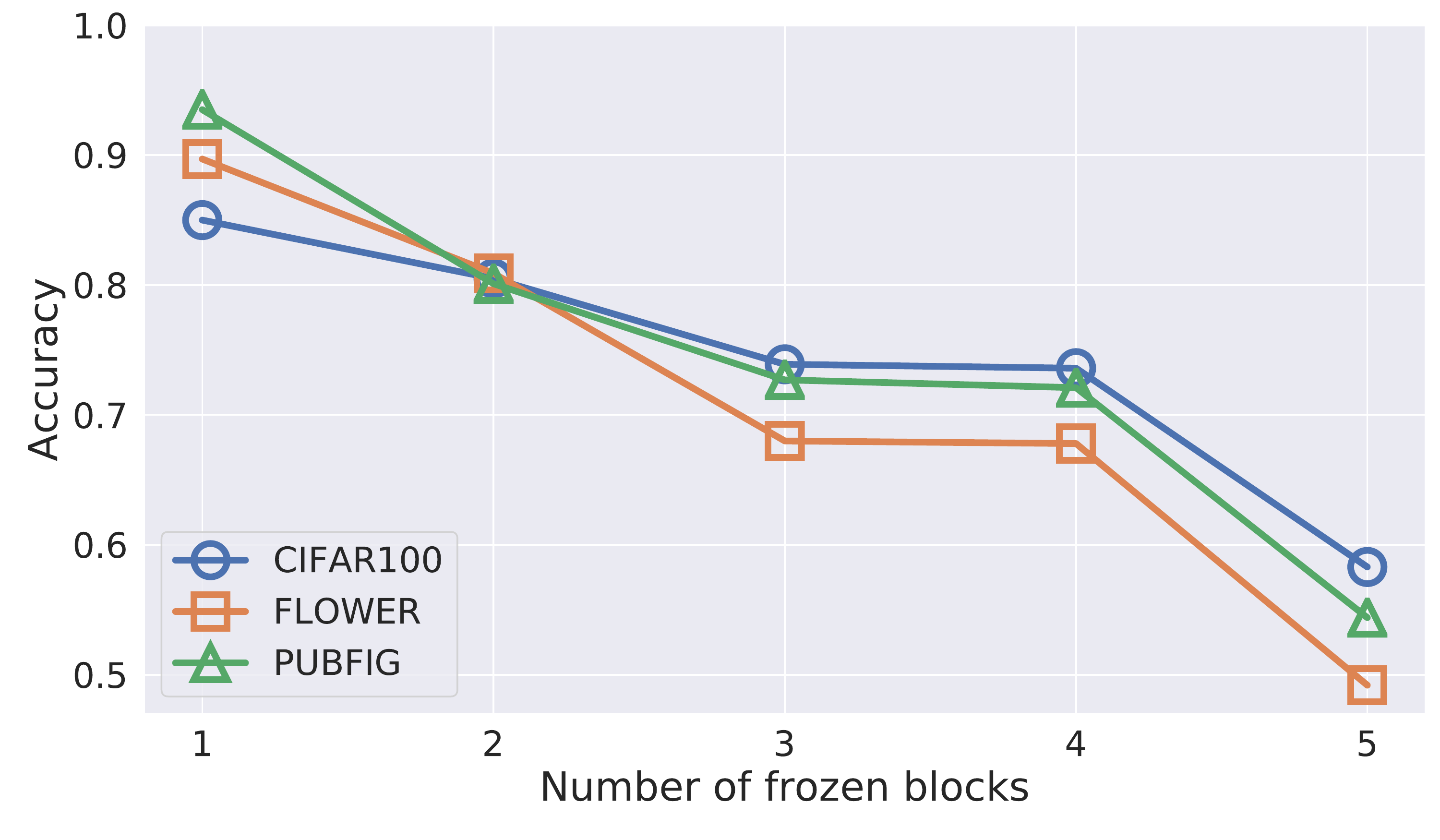}

\includegraphics[width=0.46\textwidth]{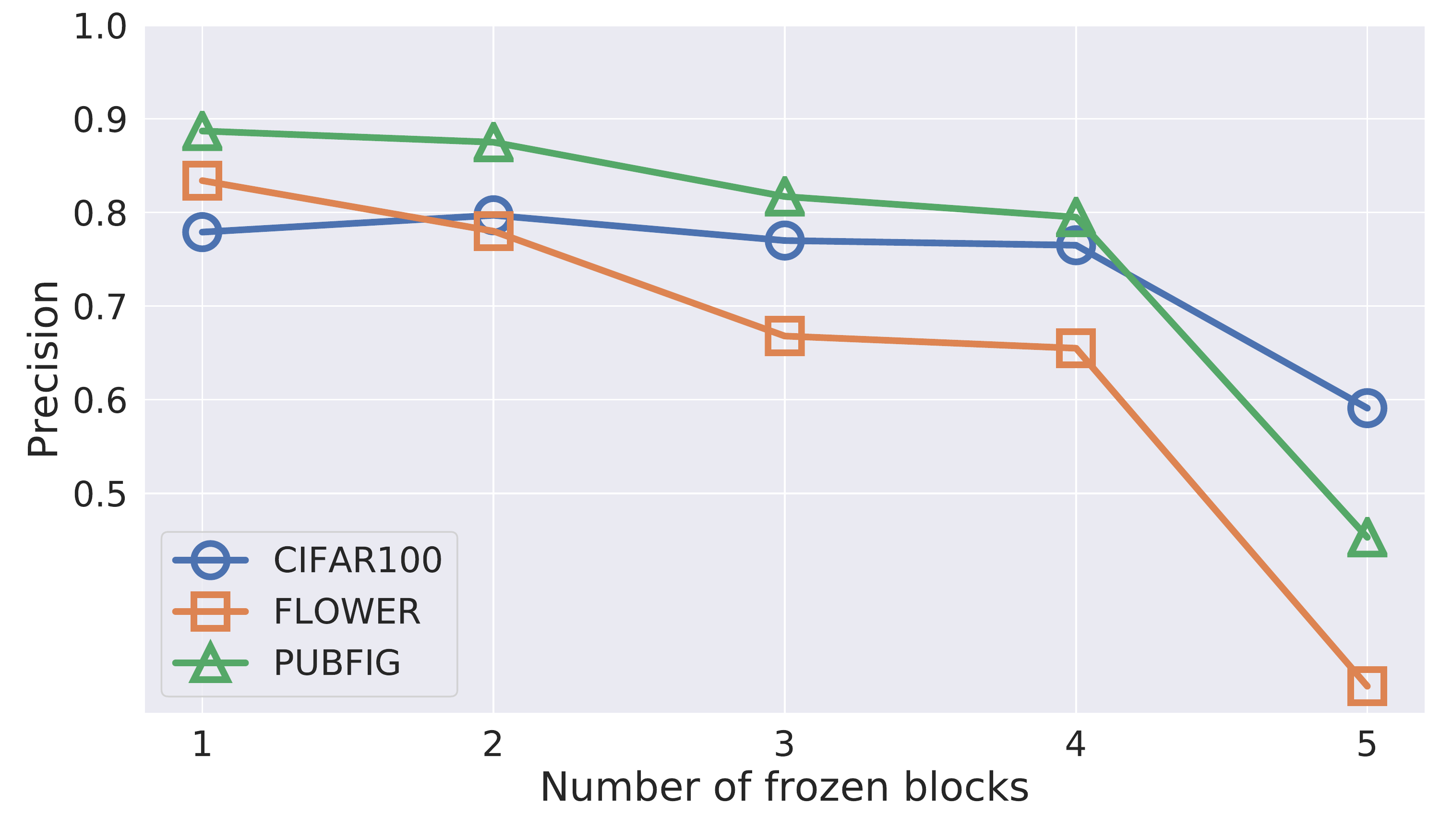}
\includegraphics[width=0.46\textwidth]{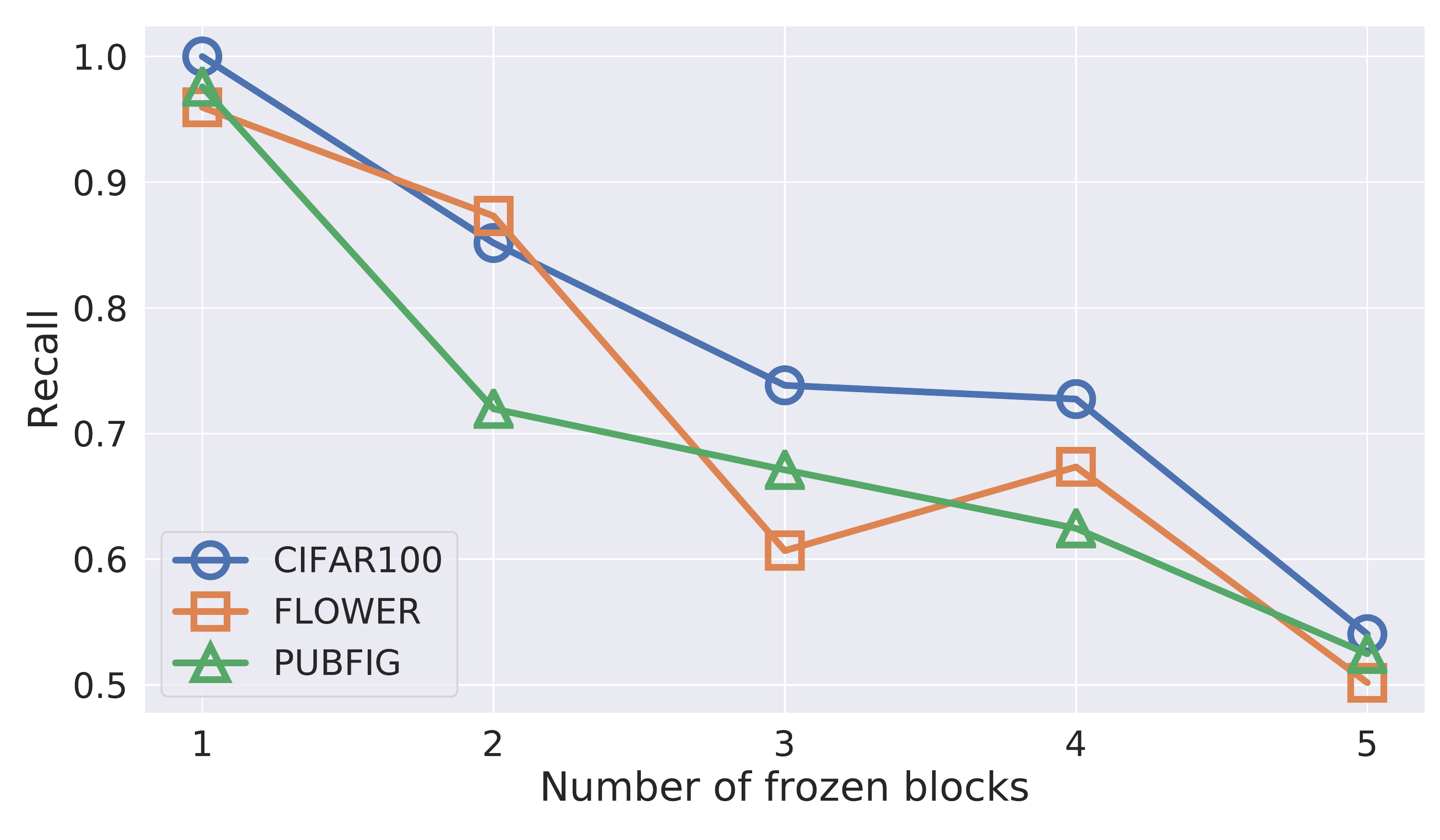}
\caption{Impact of the number of frozen blocks. 
The target model is pretrained on Caltech101 dataset and fine-tuned on three datasets to obtain three student models.
The x-axis represents the number of frozen blocks and the y-axis represents the four attack performance metrics respectively. 
The legends stand for three student datasets.}
\label{figure:metric_of_frozen_layers}
\end{figure*}

\mypara{Summaries}
Through extensive experiments, we draw the following conclusions:
1) having access to the teacher model can severely leak the membership information of teacher dataset;
2) the student model would not leak membership information of teacher dataset;
3) the student model could inevitably leak membership information of student dataset.

\subsection{Q2: Impact of Frozen Layers}
\label{section:exp2frozenlayer}

As discussed in the background part, when the ML models are complex, we always adopt the feature extractor approach to train the student model, i.e., freeze parts of the layers of teacher model and update the parameters of the last several layers using student dataset.
In this subsection, we want to know what is the impact of the number of frozen layers on the attack performance.

\begin{figure*}[!ht]
\centering  %
\includegraphics[width=0.75\textwidth]{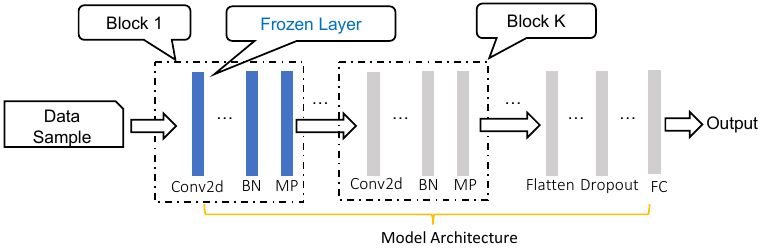}
\caption{Illustration of ``block'' frozen.
A block is defined as a set of consecutive layers separated by the max-pooling layers.
Freezing a block means fixing all the parameters of the layers within this block. 
This experiment focuses on freezing blocks from first layer to last layer.
The range of the number of frozen block is $K\in[1,5]$}
\label{figure:overview_of_frozen_layer}
\end{figure*}

\mypara{Setup}
Recall that we use ResNet20 in our experiment, which consists of $5$ blocks, each contains two convolutional layers, one batch normalization layer and one max-pooling layer.
In this experiment we freeze blocks instead of freezing each individual layer, as illustrated in Figure~\ref{figure:overview_of_frozen_layer}.
We freeze the continuous $K$ blocks on the model from first layer to last layer, and $K$ is in the range of $[1, 5]$.
In this experiment, we focus on Attack-3, and the experimental results are shown in Figure~\ref{figure:metric_of_frozen_layers}.

\mypara{Observations}
The experimental results show that the attack performance consistently drops when the number of frozen layers increase.
In the deep learning model, the higher the layer, the more abstract the features.
When we freeze the higher layers, it is more difficult for the student model to remember the student training data; thus the attack performance is worse.

When the frozen layers increase from $4$ to $5$, the attack performance drops significantly.
The reason is that there are no parameters to be updated and the student model is exactly the same with the teacher model; 
thus, the student datsets have no impact on the target models.

\begin{table}[!tpb]
 \centering
 \caption{Attack performance of transfer attack.
 The shadow models and target models are trained on VGG16 and ResNet20, respectively.
 Both of them are first pretrained on Caltech101 dataset and fine-tuned on three different datasets.
}
 \label{table:result_of_transfer_attack}
 \begin{tabular}{l l c}
  \toprule
    Shadow$\rightarrow${Target} & Dataset & AUC \\ 
  \midrule
    \multirow{3}*{\textbf{VGG16$\rightarrow${ResNet20}}} & CIFAR100 & 0.947 \\
    ~ & Flowers102 & 0.964 \\
    ~ & PubFig83 & 0.977 \\
  \bottomrule
 \end{tabular}
\end{table}

\subsection{Q3: Effectiveness of Transfer Attack}
\label{section:transfer_attack}
So far, we assume the model architecture of shadow model and target model are the same, e.g., both of them are ResNet20.
However, in practice, we only have black box access to the teacher model or student model and could not obtain the architecture of the target model easily.
Thus, in this subsection, we evaluate the attack performance when the architecture of shadow model and target model are different.
In another word, we aim to investigate the transferability between shadow model and target model.

\mypara{Setup}
We use VGG16 and ResNet20 to build the shadow model and target model, respectively.
We conduct the experiments on three dataset, i.e., CIFAR100, Flower102, and PubFig83.
We use the fine-tuning approach for transfer learning and focus on Attack-3.
The experimental results are given in Table~\ref{table:result_of_transfer_attack}. 

\mypara{Observations}
We observe that the attack AUC of all datasets are larger than $0.9$, and only drops by at most $3\%$ compared to Attack-3 in Q1.
The experimental results show that the membership inference is still effective when the architecture of the target models are unknown.

\section{Conclusion}
\label{section:conclusion}

This paper performs the first systematic study of membership inference attack against transfer learning models.
We have discovered two different attack surfaces in the transfer learning setting with two adversarial capabilities, resulting in three different attacks.
We systematically design the attack pipeline and empirically evaluate the attack performance of the three attacks on four real-world datasets.

The experimental results show that membership inference attacks against teacher and student model are both effective.
However, when the model is transferred from teacher model to student model, the membership information of teacher model will not be leaked.
We further reveal that freezing more layers during training the student model leads to worse attack performance.
Finally, we show that the membership inference attack is still effective when the architecture of the target model is unknown to to the adversary.
We believe that our studies can lead to future works on designing transfer learning in a privacy-preserving way.

\bibliographystyle{plain}
\bibliography{normal_generated_py3.bib}
\end{document}